\documentclass[aps,prb,superscriptaddress,twocolumn,showpacs]{revtex4}

\usepackage{graphicx}
\usepackage{amsmath}

\usepackage{epsfig}
\usepackage{color}
\usepackage{subfigure}

\begin{document}

\title{Quantized invariant tori in Andreev billiards of mixed phase space}

\author{Z.~Kaufmann}
\affiliation{Department of Physics of Complex Systems, E{\"o}tv{\"o}s
University, H-1117 Budapest, P\'azm\'any P{\'e}ter s{\'e}t\'any 1/A,
Hungary}
\author{A.~Korm\'anyos}
\affiliation{Department of Physics, Lancaster University, Lancaster,
LA1 4YB, UK}
\author{J.~Cserti}
\affiliation{Department of Physics of Complex Systems, E{\"o}tv{\"o}s
University, H-1117 Budapest, P\'azm\'any P{\'e}ter s{\'e}t\'any 1/A, Hungary}
\author{C.~J.~Lambert}
\affiliation{Department of Physics, Lancaster University, Lancaster,
LA1 4YB, UK}

\begin{abstract}

Comparing the results of  exact quantum calculations and  those obtained
from the EBK-like quantization scheme 
of Silvestrov \emph{et al} [Phys.~Rev.~Lett. \textbf{90}, 116801 (2003)]
we show that the spectrum 
of  Andreev billiards of mixed phase space 
can basically be decomposed into a regular and an irregular part, 
similarly to normal billiards. We provide  the first numerical 
confirmation of the validity of this quantization scheme 
for individual eigenstates
and   discuss its accuracy.

\end{abstract}


\maketitle

For quantum systems,  whose classical analogues have mixed  dynamics,
the separation of the spectra into regular and irregular parts
goes back to Percival\cite{ref:percival}. Here the term 'regular' refers to 
such energy levels, which correspond to quantised invariant tori,
while 'irregular' refers to those  associated with the chaotic part of 
the phase space. Semiclassical methods, such as  EBK quantization\cite{ref:EBK},
have given a deep understanding of the influence of the
classical dynamics on the quantum spectrum 
(see eg \cite{ref:berry-leshouche,ref:bohigas,ref:ketzmerick} 
and references therein).

Recently a new type of  quantum dot system,
comprising normal-superconductor (N-S) interface has attracted considerable attention.
These systems are commonly called "Andreev-billiards" (ABs)
\cite{ref:andreev-bil,ref:beenakker-rev}. The name derives from the specific
scattering process taking place at the N-S interface, namely  
the Andreev-reflection\cite{ref:andreev}, 
whereby an impinging electron-like quasiparticle
with energy $\varepsilon$ (measured from the Fermi energy $E_{\rm F}$) 
is coherently scattered into 
a hole (and vice versa) if  $\varepsilon$ is smaller than the superconducting gap $\Delta $
(for details see eg \cite{ref:beenakker-rev}). 
Unlike  specular reflection, this scattering process is accompanied by the 
(approximate) reversal 
of all velocity components, thereby giving rise to a peculiar classical dynamics.
In a recent paper\cite{ref:SA_wavefunc} we have presented evidence that in general  
the phase space of these systems is mixed. Thus the question naturally arises, 
of whether one can perform a similar separation of the spectrum into 
regular and irregular parts as in normal systems. 
While the influence of  different regions of the mixed phase space of an
\emph{isolated} normal dot  
on the density 
of states of the corresponding ABs has been adressed beforehand\cite{ref:schomerus} 
to our knowledge no study  has tried to answer the above  question 
regarding the individual energy 
levels and taking into account the peculiar dynamics of the \emph{whole} N-S system.  
Our work is aimed to be the first step towards the answer 
by studying a simple yet nontrivial 2D example.

In our study we shall use semiclassical and quantum 
mechanical tools  to  identify  the regular eigenstates. 
Quantum mechanically, these systems can be described by the 
Bogoliubov--de Gennes equations~\cite{ref:BdG}. 
From semiclassical point of view, this
implies that the EBK quantization has to be generalized to the case of 
spinor wave function. 
Such a  multicomponent semiclassical theory for the N-S 
systems has been developed in Ref.~\cite{ref:gyorffy} and  it was shown that  
the EBK quantization of normal  systems  can also be  generalized 
to integrable N-S systems. Indeed, in  Ref.~\cite{ref:florian} a good agreement 
between semiclassical and quantum calculations has been found for an integrable AB.
Another important idea regarding the quantisation of N-S systems was introduced 
in Refs.~\cite{ref:silvestrov,ref:jacquod} for cases where for $\varepsilon>0$
the classical dynamics is strictly speaking non-integrable, but
there is an adiabatic invariant.
At zero magnetic field the invariant is the 
time $T(\varepsilon)$ between subsequent Andreev reflections.
According to this approach, for the purpose of semiclassical
quantization, one can consider the
curves of constant $T$ at $\varepsilon=0$.

In our work we benefit from both of the studies\cite{ref:gyorffy,ref:silvestrov}.
As it will be shown
below, the dynamics in certain regions of the phase space is quasi-integrable. 
By quasi-integrable we mean regions which contain mostly tori, on which the
dynamics is similar to that of integrable normal systems. 
Therefore we expect that 
for these islands of regular motion the 
results of Ref.~\cite{ref:gyorffy} apply.  
However, the analytical calculations can be performed only in the adiabatic 
approximation\cite{ref:silvestrov}, whose accuracy 
will also be  discussed.

\begin{figure}[hbt]
\includegraphics[scale=0.4]{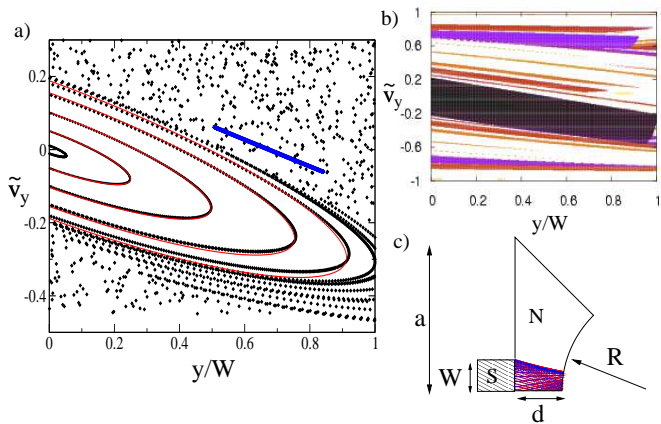}
\vspace*{-4mm}
\caption{ (colour online) Enlargement of a part of the PS around 
$\tilde{v}_y=0$ (a) (for the geometrical and other 
parameters see \cite{ref:geom-param}).
Each dot represents a starting point 
of an electron trajectory. The red curves show the ellipse-approximation 
given by Eq.~(\ref{eq:adiabatic-tori}), for different $T_0$s. The blue
curve denotes an another, thin regular island. 
An alternative view of PS showing bands of approximately constant 
$T_{eh}$ denoted by different colours (b). 
Only bands with $T_{eh}<6 a/v_{\rm F}$ are shown.
One of the tori shown in (a) is projected onto the real space (c). 
The electron trajectories are blue, the hole ones are red. 
\label{fig:poinc-integ}}
\end{figure}

The model system we used in our calculations  is the 
Sinai-Andreev (SA) billiard\cite{ref:effective_rmt,ref:SA_wavefunc}. 
It consists  of a Sinai-billiard-shaped 
normal dot and an attached (infinite) superconducting 
lead as shown in Fig.~\ref{fig:poinc-integ}(c). 
Classically, in the phase space of  this model we have found  
a large stability island (see Fig.~\ref{fig:poinc-integ}(a)).  
Its existence can be expected, since
it is centered on the shortest unstable periodic orbit of the isolated normal dot 
(which corresponds to the motion along the bottom wall) and the
presence of the superconductor, i.e.,  Andreev-reflection can 
stabilise this orbit. 
To prove this, we consider the dynamics on the Poincar{\'e} section (PS) 
which we define in the
following way: we record the position $y$ and the tangential 
velocity component $v_y$ in units of 
$v_e =v_F\sqrt{1+\tilde{\varepsilon}}$ 
when the Andreev reflection results in a departing electron quasiparticle, ie whenever a 
hole impinges on the N-S interface   
(here   $\tilde{\varepsilon}=\varepsilon/E_{\rm F}$, $v_F$ is the Fermi velocity
and $y$ is measured from the lower edge of the interface). 
Suppose now that an electron 
departs from the interface and after returning to it becomes Andreev-reflected.  
Using simple geometrical considerations and taking into account the Andreev
reflection law\cite{ref:shytov,ref:adagideli}  one finds that for 
$y\ll 1$, $v_y\ll v_e$
the linearised equations of motion for the phase space coordinates
$y$, $\tilde{v}_y=v_y/v_e$
of the quasiparticle is given by the stability matrix 
\begin{equation}
M(\varepsilon)=
\left(\begin{array}{cc} 
1-2 d/R  & 2d(1-d/R) \\ 
\gamma \, 2/R & \gamma\,(2 d/R-1) 
\end{array}\right)
\label{eq:electron-map}
\end{equation}
where 
$\gamma=\sqrt{1+\tilde{\varepsilon}}/ \sqrt{1-\tilde{\varepsilon}}$ and the 
geometrical parameters $d$, $R$ are defined in Fig.~\ref{fig:poinc-integ}(c).   
The motion of the emerging hole  can then be described by $M(-\varepsilon)$. 
However, by construction the Poincar{\'e} map consists only of  the starting coordinates  of
electron trajectories,  therefore the motion in the $(y,\tilde{v}_y)$ plane is given by
$M_{eh}=M(-\varepsilon)\,M(\varepsilon)$. The periodic orbit is stable, if 
the trace of  the matrix $M_{eh}$ is less  than $2$, which gives the 
condition $0<d< R/(\gamma-1)$. 
Since  $\varepsilon\ll E_{\rm F}$ 
in our calculations, $1/(\gamma-1)\approx 1/ \tilde{\varepsilon} \gg 1$ and  
the periodic orbit is stable 
for wide range of  values of the  parameters $d$ and $R$.

Part of the PS around the stable periodic orbit is shown in 
Fig.~\ref{fig:poinc-integ}(a). 
\vspace*{-2mm}
As expected, we find numerically that there are truly 
invariant curves around the stable periodic
orbit forming an island of regular motion. This regular island, which can be
clearly seen in the middle of  Fig.~\ref{fig:poinc-integ}(a) is
surrounded by intermittent-like regions.
Beyond  there are chaotic seas intervowen  with other intermittent
regions and there is at least one another smaller regular island.
As  pointed out in our previous paper\cite{ref:SA_wavefunc}, 
this structure is due to the interplay of {\em non-exact retracing} 
for $\tilde{\varepsilon}>0$ and the presence of the 
so called "critical points" which separate the normal 
and the superconducting segments of the billiard 
boundary.

Because of its importance in the semiclassical 
quantization, we now focus on the energy dependence of the time 
$T_{eh}(y, \tilde{v}_y, \tilde{\varepsilon})$
of the electron-hole orbit between two section with the PS in the regular island. 
Following the electron-hole orbits which trace out 
the invariant curves $\mathcal{C}_{inv}$ 
in Fig.~\ref{fig:poinc-integ}(a), we have found numerically that
(a) the variation of $T_{eh}(y, \tilde{v}_y, \tilde{\varepsilon})$ 
corresponding to subsequent points on the  invariant curves
is only $\mathcal{O}(\tilde{\varepsilon}^3)$, 
ie  $T_{eh}$ is an adiabatic invariant
(b) 
$T_{eh}=2T(y, \tilde{v}_y,\tilde{\varepsilon}=0)+\mathcal{O}(\tilde{\varepsilon}^2)$,
where we denoted
by $T$ the time between \emph{subsequent} Andreev-reflections for 
$\tilde{\varepsilon}=0$.
An important consequence of the above 
observations is the following: if we denote by $\mathcal{C}_{T_0}$ the 
$T(y, \tilde{v}_y,\tilde{\varepsilon}=0)=T_0$ curves on the PS, then
considering any point $(y, \tilde{v}_y)$ on $\mathcal{C}_{T_0}$,  
for finite  $\tilde{\varepsilon}$ the 
invariant curve  $\mathcal{C}_{inv}$ 
which contains this particular point 
will always be in  $\mathcal{O}(\varepsilon^2)$ vicinity of 
$\mathcal{C}_{T_0}$. 
This will be important later on when we discuss the semiclassical quantization
of this regular island: it can be 
shown\cite{ref:gyorffy,ref:silvestrov},  that 
one of the action integrals to be calculated 
equals the area enclosed by $\mathcal{C}_{inv}$, which can be then approximated 
by the area enclosed by $\mathcal{C}_{T_0}$, since the difference is of
higher   order than $\mathcal{\tilde{\varepsilon}}\ll 1$. 
A similar conclusion has
been drawn in Ref.\cite{ref:silvestrov}.

An alternative  view of the PS can be obtained by denoting with
different colours those regions, which correspond to approximately 
constant  $T_{eh}$.
These regions, which usually appear as narrow bands 
in Fig.~\ref{fig:poinc-integ}(b) are often not easily 
recognizable on the  PS 
since they do not always show ordered pattern such as  those
indicating the presence of intermittent-like motion 
around the quasi-integrable island in Fig.~\ref{fig:poinc-integ}(a). (The 
large dark region
in the middle of  Fig.~\ref{fig:poinc-integ}(b) corresponds to the regular 
island and the surrounding intermittent region 
in Fig.~\ref{fig:poinc-integ}(a).) As we will  briefly discuss  later,
the importance of these bands  lies in the fact that   
some of them  can be associated to certain eigenstates. 

Before turning to the adiabatic quantization, 
we note that for a different $W$,  and larger value of  
$\tilde{\varepsilon}$, 
in the regular island we have also found  structures 
resembling very much the secondary island chains known from 
normal KAM systems. It would be an interesting future project to
investigate  the properties of the classical dynamics in more detail 
to explore the exact nature of  the similarities  
with KAM systems.
However, for the parameters\cite{ref:geom-param} used in  the 
present study, these "secondary islands", if they exist,
are much smaller than Planck's constant $h$ and thus  do not
play any role in our further discussion.

We now proceed with the adiabatic quantisation\cite{ref:silvestrov} 
of the regular island shown in Fig~\ref{fig:poinc-integ}(a). 
The semiclassical energies can be obtained by quantising the 
action integrals $I_i=1/2\pi \oint_{\mathcal{C}_i}\mathbf{p}\,{\rm d}\mathbf{q}$, $i=1,2$, 
calculated along the $\mathcal{C}_i$ irreducible closed contours on the adiabatic tori. 
The curve $\mathcal{C}_{T_0}$  
can be chosen as the integration contour  $\mathcal{C}_1$ and therefore
$I_1$ equals  the enclosed area on the PS. 
One can show  that for   $y, \tilde{v}_y\ll 1$  the curve $\mathcal{C}_{T_0}$ is 
semi-ellipse given by the following equation:
\begin{equation}
(L-2 d)R^2=F\, y^2+G\,\tilde{v}_y\,y+H\, \tilde{v}_y^2
\label{eq:adiabatic-tori}
\end{equation} 
where $F=(2d+R)$, $G=4d(d+R)$, $H=F G/4$ and $L=v_{\rm F} T_0$. 
Comparison of the above curve for different $T_0$s and the numerically-calculated PS 
of the corresponding tori for  
$\tilde{\varepsilon}=0.0105$ 
is shown in Fig.~\ref{fig:poinc-integ}(a). One can observe that the agreement is very good
for the inner tori, while  for those closer to the border of the regular island there is a 
small but noticeable deviation since the assumption $y, \tilde{v}_y\ll 1$ does not hold. 
Quantising  the area enclosed by $\mathcal{C}_{T_0}$ one finds  the condition
\begin{equation}
\frac{1}{8}\frac{(L-2d)R}{\sqrt{d(d+R)}}\,p_{\rm F}=\hbar\left(m+\frac{3}{4}\right)
\label{eq:quant-cond-1}
\end{equation}
where $m=0, 1, 2,\dots$ and
the Maslov index is $3/4$ since  one caustic and one hard wall is encountered 
along  the integration contour.

The second action integral $I_2$ in the adiabatic approximation is
to be calculated along a self-retracing electron-hole orbit. 
Since  the particles move ballistically inside the normal dot, 
the quantisation condition reads:
\begin{equation}
(p_e-p_h)L=2\pi\hbar[n+\frac{1}{\pi}\arccos(\varepsilon / \Delta)].
\label{eq:quant-cond-2}
\end{equation} 
Here $p_e(\tilde{\varepsilon})=p_{\rm F}\sqrt{1+\tilde{\varepsilon}}$ and 
$p_h(\tilde{\varepsilon})=p_e(-\tilde{\varepsilon})$ are the magnitudes 
of the electron's and hole's momentum respectively, 
$n=0,1,2,\ldots$, and with the 
$\arccos(\varepsilon / \Delta)$ term we take into account the phase shift due to the 
Andreev-reflection. Combining Eqs.~(\ref{eq:quant-cond-1}) and (\ref{eq:quant-cond-2}) 
one  arrives at an implicit equation for the eigenvalues $\varepsilon_{nm}$ 
which can be  solved  numerically for each value of $n$ and $m$.

Before comparing  the semiclassical and exact quantum calculations,
we briefly discuss the accuracy of the adiabatic approximation.
There are two sources of error: the first  is  in the derivation of 
Eqs.~(\ref{eq:quant-cond-1}) and (\ref{eq:quant-cond-2}) where 
we considered the invariant
surfaces in the phase space for $\tilde{\varepsilon}=0$. However, 
due to the non-exact retracing,
the dynamics is  different for finite $\tilde{\varepsilon}$. Since  
in the quantum  system the excitation energies $\tilde{\varepsilon}_{nm}$ 
are always finite,  one should  calculate the action integral $I_1$ 
using the invariant curves $\mathcal{C}_{inv}$, 
whereas in case of  $I_2$ one should take into account that 
for $\tilde{\varepsilon}>0$ the hole does not retrace exactly the path 
of the electron, and one should choose the integration contour accordingly.  
It turns out however, that for this regular island the results for $I_1$ and $I_2$ do
not change in first order of $\tilde{\varepsilon}$ even 
if we take into account the effects of 
finite  $\tilde{\varepsilon}$ on classical dynamics. 
Namely, as we have already pointed out, the difference between the areas 
enclosed by a $\mathcal{C}_{inv}$ and a corresponding
$\mathcal{C}_{T_{0}}$ is only of   
$\mathcal{O}(\tilde{\varepsilon}^2)$ and therefore $I_1$ 
is accurate to first order in $\tilde{\varepsilon}$.
Considering  $I_2$, we checked the accuracy of 
Eq.~(\ref{eq:quant-cond-2}) by taking as an    
integration contour  $\mathcal{C}_2$ 
a non-retracing electron-hole orbit and then we closed the contour on the PS. 
After lengthy calculations we have found that the
first correction to the result given by  Eq.~(\ref{eq:quant-cond-2}) 
is of the  order of $\tilde{\varepsilon}^3$.   

\begin{figure}[hbt]
\includegraphics[scale=0.4]{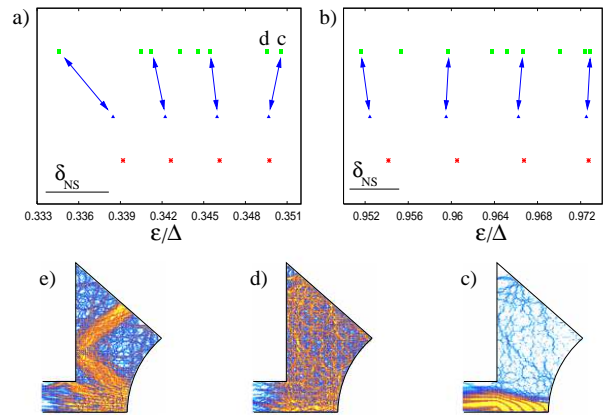}
\vspace*{-6.2mm}
\caption{(colour online) Quantum eigenenergies (green squares), the results of the 
adiabatic quantization  using Eq.~(\ref{eq:quant-cond-1}) (red stars) 
and by taking into account the exact shape of $\mathcal{C}_{T_0}$
(blue triangles) for $n=0$ (a) and $n=1$ (b).
The arrows show  
which quantum  and semiclassical energies correspond to each other. 
The bars show the 
magnitude of the mean level spacing $\delta_{\rm NS}=\delta_{\rm N}/2$. 
The square modulus $|u(\mathbf{r})|^2$ of the electron components of the wave functions 
corresponding to certain eigenenergies as indicated by the
letters c, d in Fig.~\ref{fig:qm-semi-levels}(a)
are seen in (c),(d).
The  $|u(\mathbf{r})|^2$ for the state at $\varepsilon/\Delta=0.3202$
corresponding to a band of approximately constant\-
$T_{eh}$ starting
at $y=0, \tilde{v}_y=\pm 0.4$
in Fig.~\ref{fig:poinc-integ}(b)
is shown in (e).
\label{fig:qm-semi-levels}}
\end{figure}

The second source of error is that  in order to 
calculate $I_1$ we approximated the curves 
$\mathcal{C}_{T_{0}}$ by the semi-ellipses given by 
Eq.~(\ref{eq:adiabatic-tori}). This 
certainly introduces inaccuracy   in case of adiabatic tori close to 
the border of the regular island, for which 
the conditions $y, \tilde{v}_y\ll 1$ do not rigorously hold.
Therefore we checked our analytical results in the following way: 
first we numerically determined those $\mathcal{C}_{T_{0}}$ curves
on PS for which the enclosed area is $\hbar(m+3/4)$, $m=0,1,...$. 
Then reading off the $T_0$ values corresponding to these curves and using
Eq.~(\ref{eq:quant-cond-2}) we re-calculated the semiclasssical energies. The
results are summarized in Fig.~\ref{fig:qm-semi-levels}(a),(b) and will be discussed 
below. 

We now  compare  the results of the quantum mechanical and semiclassical
calculations and show that the semiclassical energies  agree remarkably well
with the exact quantum ones.
The quantum treatment of the system 
is based on the Bogoliubov--de Gennes equations\cite{ref:BdG} 
and is briefly described in \cite{ref:SA_wavefunc}. 
It is  assumed  that the  superconducting pair potential  $\Delta$ is 
constant inside the lead and zero in the N region\cite{ref:beenakker-rev} 
and we work in the regime 
$\delta_N\ll E_{\rm T}\ll \Delta\ll E_{\rm F}$\cite{ref:beenakker-rev} 
where $\delta_N$ is the mean level spacing of the isolated normal dot and $E_{\rm T}$ is 
the Thouless energy\cite{ref:geom-param}.
The area of the regular island on the PS is $\approx 4 h$ and therefore we expected
$8$ regular eigenstates in the spectrum, corresponding to quantum numbers
$n=0,1$, $m=0,1,2,3$ (For the parameters\cite{ref:geom-param} used in our calculation
Eqs.~(\ref{eq:quant-cond-1}), (\ref{eq:quant-cond-2}) do not have real solution for 
larger $n$, while the  values of $m$ are limited by the size of the stability
island). 
\vspace*{-2mm}
\vspace*{-2mm}

Comparison of the semiclassical predictions for $n=0$, $m=1\dots 4$  and the
exact quantum mechanical eigenvalues lying roughly in the same energy range 
can be seen in Fig.~\ref{fig:qm-semi-levels}(a) and (b). 
This shows that there are twice as many quantum eigenenergies as 
semiclassical ones and at first sight it is not obvious  which of the eigenstates 
should be considered as regular ones. 
To identify the eigenstates which correspond
to quantized tori we now examine the eigenfunctions. 
Similarly to normal billiards\cite{ref:robnik} we have found that 
the wave function 
of certain eigenstates show  strong  localization (both the electron
and the hole components)  
onto classical objects (see also \cite{ref:SA_wavefunc,ref:florian}).  
This can be either a torus  
(compare Fig.~\ref{fig:qm-semi-levels} (c) and
Fig.~\ref{fig:poinc-integ}(c)),  
or a bunch of trajectories (see eg Fig.~\ref{fig:qm-semi-levels} (e)) 
corresponding to a band of approximately 
constant $T_{eh}$ in Fig.~\ref{fig:poinc-integ}(b).

Computing  eg the eigenfunctions  belonging to the eigenvalues shown in 
Fig.~\ref{fig:qm-semi-levels}(a) reveals that 
only four of them are localized (as in Fig.~\ref{fig:qm-semi-levels}(c)), 
three others display an apparently random interference 
pattern and cover  the normal dot in roughly uniform way
(as in Fig.~\ref{fig:qm-semi-levels}(d)), 
while one of them is of intermediate nature.
(The picture is very similar in  case of Fig.~\ref{fig:qm-semi-levels}(b)).
As an example, we show the electron component of
those  two eigenstates  energies of which are very close to the $n=0,m=0$ 
semiclassical one (see Figs.~\ref{fig:qm-semi-levels}(c), (d)). 
One can clearly see that 
one of them is chaotic, while the other  is localized onto a classical torus. 
Thus the criteria for accepting that a quantum eigenstate  
corresponds to a quantized 
torus are the vicinity of the eigenenergy to the semiclassical prediction
\emph{and} localization of the eigenfunction.
Based on these  two criteria, we indeed identified $4+4$  regular eigenstates
(for the $n=0,1$, $m=0\dots3$ cases) 
in the spectrum. 
The accuracy of the adiabatic quantization is remarkable, 
since using the numerically-determined 
$\mathcal{C}_{T_{0}}$ for the semiclassical quantization shows that
the difference between the 
quantum and semiclassical energies is $\sim 10^{-1} \delta_{NS}$,
except 
for the $n=0$, $m=3$ eigenstate [the leftmost one in 
Fig.~\ref{fig:qm-semi-levels}(a)], for which it is $0.83\delta_{NS}$
(here $\delta_{NS}=\delta_{N}/2$ is the mean level spacing of the N-S system).
If one approximates the  curves  $\mathcal{C}_{T_{0}}$ by semi-ellipses as 
in Eq.~(\ref{eq:quant-cond-1}),
the agreement remains the same for the states with quantum number $m=0,1$ 
and slightly deteriorates for the $m=2,3$ states.       
Finally,
the observation that chaotic eigenstates are intermixed with regular ones
in the given energy interval        
suggests that the Berry-Robnik conjecture\cite{ref:berry-robnik} 
for the spectrum of  (normal) systems with mixed classical dynamics  might
also  hold for ABs. 

Besides the regular eigenstates discussed so far, we have found 
that some of the bands of nearly constant $T_{eh}$ shown 
in Fig.~\ref{fig:poinc-integ}(b) with intermittent dynamics
also support one or more quantum eigenstates.
The energies of these  eigenstates can also be obtained with semiclassical
methods, although not as accurately as in the previous case. 
We found that calculating the average time $\overline{T}_{eh}$ of 
an electron-hole orbit in a given band and then    
using Eq.~(\ref{eq:quant-cond-2}) with 
$\overline{L}=v_{\rm F}\overline{T}_{eh}$ one can usually predict the 
quantum eigenvalues
with an error $ \lesssim \delta_{NS}$. As an example, 
for the eigenstate shown in
Fig.~\ref{fig:qm-semi-levels}(e) the quantum calculation gives  
$\varepsilon/\Delta=0.3202$ 
while for $n=1$ the semiclassical result is   $0.3239$, 
giving an  error of $0.8\delta_{NS}$. (There is also a quantum state which 
corresponds to the $n=0$ semiclassical one, but it is not as 
clearly localised as the one shown in Fig.\ref{fig:qm-semi-levels}(e).)

In summary, semiclassical analysis and 
the wave function computation enable the classification of  
certain  eigenstates as  regular ones 
for Andreev billiards of mixed phase space.
For regular states we present the first numerical calculation
to show that  EBK-like quantization scheme yields good agreement with the
quantum results. Moreover, other  states are either chaotic or can be 
associated with 
bands on the PS for which the time 
until the next Poincar\'e 
section is approximately constant. 

We would like to thank H. Schomerus 
for useful discussions.
This work is supported
by E. C.
Contract No. MRTN-CT-2003-504574, EPSRC,
the Hungarian-British TeT.


\begin{thebibliography}{99}
\bibitem{ref:geom-param}
Throughout this paper the number of open channels in the lead is the integer part 
of $k_{\rm F} W/ \pi=25.51$ ($k_{\rm F}$ is the Fermi wavenumber), while 
$\Delta_0/E_{\rm F}=0.03$ and $W=0.21$, $R=0.7$ in units of $a$. For   
Fig.~\ref{fig:poinc-integ}(a) we used   
$\tilde{\varepsilon}=0.0105$ in the actual calculation.
\end{thebibliography}
\end{document}